\documentclass[]{pasj01} % 投稿時はproofにする

\Received{$\langle$reception date$\rangle$}
\Accepted{$\langle$acception date$\rangle$}
\Published{$\langle$publication date$\rangle$}
\usepackage[switch,mathlines]{lineno}
\usepackage{ulem}

\begin{document}

\title{Cosmic very small dust grains as a natural laboratory of mesoscopic physics: Modeling thermal and optical properties of graphite grains}

\author{Kenji \textsc{Amazaki}\altaffilmark{1,2,}\footnotemark[*]}
\author{Masashi \textsc{Nashimoto}\altaffilmark{3},}
\author{Makoto \textsc{Hattori}\altaffilmark{1}}

\altaffiltext{1}{Astronomical Institute, Tohoku University, Aza-Aoba 6-3, Aoba-ku, Sendai, Miyagi 980-8578, Japan}
\altaffiltext{2}{Graduate Program on Physics for the Universe (GP-PU), Tohoku University, Aza-Aoba 6-3, Aoba-ku, Sendai, Miyagi 980-8578, Japan}
\altaffiltext{3}{Graduate School of Science, The University of Tokyo, 7-3-1 Hongo, Bunkyo-ku, Tokyo 113-0033, Japan}
\email{kenji.amazaki@astr.tohoku.ac.jp}


\maketitle

\begin{abstract}
Cosmic very small dust grains (VSGs) contain 100 to 10,000 atoms, making it a mesoscopic system with specific thermal and optical characteristics due to the finite number of atoms within each grain. This paper focuses on graphite VSGs which contain free electrons. 
The energy level statistics devised by Kubo (1962, J.Phys.Soc.Jpn., 17, 975-986) were used for the first time to understand the thermal properties of free electrons in graphite VSGs.
We showed that the shape irregularity of the grains allows graphite VSGs to absorb or emit photons at sub-millimeter wavelengths or longer; otherwise, the frequency is limited to above a few THz. 
Additionally, we considered the decrease in Debye temperature due to the surface effect. 
VSGs have an extremely small volume, resulting in limited thermal energy storage, especially at low temperatures. 
Since a VSG is able to emit a photon with energy smaller than its internal energy, this determines the maximum frequency of the emitted photon.
We developed a Monte-Carlo simulation code to track the thermal history of a dust grain, considering the stochastic heating from the absorption of ambient photons and radiative cooling. 
This approach was applied to the interstellar environment to compute the spectral energy distributions from the interstellar graphite dust grains.
The results showed that graphite VSGs emit not only the mid-infrared excess emission, but also a surplus emission from sub-millimeter to millimeter wavelengths. 

\end{abstract}

\section{Introduction\label{sec:intro}}
The whole sky looked up from the Earth is covered by the Galactic interstellar dust grains (e.g., \cite{Schlegel1998, Doi2015}). 
Concrete observational evidence \citep{Weingartner2001} verifies the presence of nano-sized dust grains, commonly referred to as cosmic very small dust grains (VSGs).
These VSGs typically contain anywhere from 100 to 10,000 atoms.
This size places VSGs at a unique spot between microscopic and macroscopic dimensions, which is known as the mesoscopic scale. 
In this study, we develop a model that illustrates the thermal and optical characteristics of very small graphite dust grains incorporating the mesoscopic physics.

Graphite grains, comprised of carbon atoms, contain free electrons. 
Every carbon atom within the graphite offers four valence electrons with three engaged in creating covalent bonds.
These bonds connect carbon atoms to form a molecular panel known as graphene.
The remaining valence electron is utilized to stack graphene sheets into layers through a weak $\pi$ bond \citep{Atkins2017}. 
Free movement of $\pi$ electrons within the graphene sheet is possible, and it is believed that these electrons behave as free electrons.

The finiteness of the grain size impacts its thermal properties in two ways. 
Firstly, there is the surface effect, which reduces the Debye temperature due to VSGs having a large surface-to-volume ratio. 
As indicated in subsection \ref{sec:surf}, although this effect is noticeable, it is not overly severe. 
The other effect is the limited number of vibration modes in a VSG, which can play a role in holding thermal energy when the grain cools to a low temperature. 
This indicates that the ``temperature", which is defined by equating the internal energy of the grain $E_{\rm gr}$ to the average heat energy $\bar{E}(T)$ calculated from the statistical mechanics with the temperature $T$, loses a well-defined meaning (\cite{Draine2001}, hereafter DL01). 
The lowest energy level of the vibration mode, $\varepsilon_1$, is bounded by the grain size (DL01). 
For a spherical graphite grain, it becomes $T_1\equiv\varepsilon_1/k_{\rm B}\sim 50\,{\rm K}\, (a/1\,{\rm nm})^{-1}$ where $a$ is the radius of the spherical grain and $k_{\rm B}$ is the Boltzmann constant.
Below $T_1$, only the first excitation mode can be concerned with the radiative processes. 

The presence of the free electrons enables a considerable number of modes for thermal excitation, even within a cold VSG.
The thermal characteristics of these free electrons are effectively elucidated by the Fermi gas model \citep{Kubo1962}.
According to this model, in the ground state of the Fermi gas, all energy levels below the Fermi energy are occupied by electrons.
The thermal energy is stored by exciting electrons above the Fermi energy.
The size dependence of the number of free electrons inside a graphite grain is $N_{\rm e}= 471\, (a/1\, {\rm nm})^{3}$ (subsection \ref{sec:ele}).
The Fermi energy is given as $E_{\rm F}\sim (9\pi N_{\rm e})^{2/3}\hbar^2/(2 m_{\rm e}a^2)\sim 8$ eV, where $\hbar$ is the reduced Planck constant and $m_{\rm e}$ is the electron mass.
Therefore, the average spacing of successive energy levels of free electrons around the Fermi energy is $\delta/k_{\rm B} \sim (E_{\rm F}/N_{\rm e})/k_{\rm B}\sim 200\,{\rm K}\, (a/1\, {\rm nm})^{-3}$. 
Since the shapes of the VSGs are expected to be far from regular shapes, the degeneracy of the energy levels is broken.
Furthermore, the VSGs exhibit various irregular shapes, leading to a diverse distribution of energy levels among them.
Given the vast number of VSGs in interstellar space, the spacing between energy levels around the Fermi energy within the VSG ensemble is expected to be randomly distributed.
What we observe is emission and absorption by this ensemble of the VSGs. 
It is unnecessary to ascertain the energy level distribution of individual VSGs; rather, understanding the statistical features of the ensemble is sufficient for modeling the thermal and optical characteristics of free electrons within interstellar graphite VSGs. 
This method was initially proposed by \citet{Kubo1962} to depict the thermodynamic behavior of nano-sized metal particles, known as energy level statistics. 
A similar concept was developed in nuclear physics to model nuclei with large mass numbers \citep{Mehta1990}. 
According to the random matrix approximation, the energy level spacing around the Fermi level follows the Wigner distribution \citep{Mehta1990}. 
As the Wigner distribution allows for a continuous range of excitation energies from 0 to approximately $\delta$, applying energy level statistics significantly increases the effective number of thermal excitation modes in a cold VSG.
Consequently, introducing thermodynamic temperature may be possible for describing the thermal properties of graphite VSGs.

DL01 developed the method to calculate thermal emission from VSGs using a statistical approach. 
They computed every possible state of a grain based on a given internal energy, with the assumption that this energy was evenly distributed among all vibration modes. 
They conceptualized the radiative process as a transition between these states.
This approach circumvents the need to introduce thermodynamic temperature in calculating thermal emission from the VSGs. 
However, this calculation did not consider the contribution from the free electrons, which necessitates the development of a new method accounting for these contributions while estimating thermal emissions from graphite VSGs.

This paper introduces the application of energy level statistics to model the thermal and optical attributes of free electrons within a graphite VSG for the first time.
Additionally, it inspects how the surface effects alter the thermal properties of vibration modes.
We have developed a Monte-Carlo simulation code to track a dust grain's thermal history \citep{Draine1985}.
The code treats heating as a stochastic process and cooling as a continuous process.
Provided models describe their thermal and optical properties; this code is applicable to any kind of dust grains.
To elucidate the physical attributes of the model, the thermal emission from a graphite VSG is calculated with cosmic microwave background (CMB) radiation as the only heat source. 
The spectral energy distributions (SEDs) of the interstellar graphite from mid-infrared to millimeter wavebands predicted by our models are shown. 

The structure of this paper is as follows.
In section \ref{sec:thermprop}, we elaborate on the modeling of these properties, taking into account mesoscopic physics. 
The methodologies we employed to model thermal emissions from graphite VSGs form the subject matter of subsections \ref{sec:IETB} and \ref{sec:basemodel}.
Subsection \ref{sec:reference} presents a comparative reference model, neglecting the mesoscopic physics considerations.
A brief description of our numerical code to follow the thermal history of the dust grain is provided in subsection \ref{sec:MCcode}. In subsection \ref{sec:dustem}, we run test simulations using our numerical code to track the dust grain's thermal history and compare this data with results from the DustEM code \citep{Compiegne2011}. 
Section \ref{sec:result} is results-oriented: subsection \ref{sec:CMBonly} focuses on results when the sole heat source is the CMB, followed by a display of the probability distributions of grain temperature and SEDs from various dust grain sizes in subsection \ref{sec:CMBandstarlight}.
Section \ref{sec:discussion} is discussion-oriented: in subsection \ref{sec:flattening}, 
we demonstrate the total SEDs of thermal emission from interstellar graphite dust grains when majority of very small carbonaceous dust grains are graphite.
Subsequently, in subsection \ref{sec:limitationofIET}, we delve into the limitations of using thermodynamic temperature to characterize the thermal and optical properties of VSGs.
Further discussion on applying our approach to other dust species appears in subsection \ref{sec:extendingmodel}, supplemented by the exploration of the necessity of considering the quantum interference effect of free electrons in subsection \ref{sec:interference}.
The paper concludes with a summary and our final thoughts in section \ref{sec:conclusion}. 

\section{Modeling thermal properties of very small graphite grains}\label{sec:thermprop}

\subsection{Surface effect on thermal energy held by vibration modes \label{sec:surf}}

We examine how the surface effect alters the temperature-dependent thermal energy in vibration modes.
Atoms on a grain's surface possess lower cohesive energy than those inside.
Due to the high surface-to-volume ratio in VSGs, each atom's cohesive energy decreases compared to larger grains. 
We posit that each surface atom's cohesive energy is half of the interior atoms'. 
Thus, we can characterize a grain's total cohesive energy as follows:
\begin{equation}
 N_{\rm a}U = U_{\infty} (N_{\rm a}-N_{\rm s}) + \frac{1}{2}U_{\infty}N_{\rm s},
\end{equation}
where $U$ is the cohesive energy per atom of the grain, $U_{\infty}$ is the cohesive energy per interior atom, $N_{\rm a}$ is the total number of atoms and $N_{\rm s}$ is the number of surface atoms.
According to the Lindemann theory, the Debye temperature $\Theta$ depends on $U$ as $\Theta \propto U^{1/2}$ \citep{Reissland1973}.
Therefore, the Debye temperature of the VSG can be written as:
\begin{equation}
 \Theta = \Theta_{\infty}\sqrt{1-\frac{N_{\rm s}}{2N_{\rm a}}}, \label{eq:surfdebye}
\end{equation}
where $\Theta_{\infty}$ is the Debye temperature of the big grain. The surface effect makes the Debye temperature smaller as the grain size becomes smaller.

The total number of atoms is calculated by $N_{\rm a} = \rho VN_{\rm A}/M$ for a grain of a fixed volume $V$, where $\rho$ is the mass density of the grain, $N_{\rm A}$ is the Avogadro constant and $M$ is the molar mass. The $a$ is the effective radius of the grain defined by $a\equiv [3V/(4\pi)]^{1/3}$. 
The values of $\rho$ and $M$ for graphite are shown in table \ref{table:physparams}.
The number of surface atoms $N_s$ is calculated by dividing the effective surface area of the grain ($\eta S$) with the cross-section of the atom, where $\eta$ is the packing factor. 
The $\eta$ represents the ratio of the volume occupied by atoms and the total volume of the grain, i.e., $\eta = N_{\rm a} v / V$, where $v$ is the volume of a single atom. When the radius of the atom is $d$, assuming that atoms are spherical, the volume is $v=4\pi d^3/3$ and the cross-section is $\pi d^2$. 
Therefore, the number of the surface atoms is obtained by $N_{\rm s} = \eta S/(\pi d^2)$. 
The values of $d$ and $\eta$ are also shown in table \ref{table:physparams}. 
The surface area is obtained by assuming that the shape of the grains is ellipsoid. The surface area of an ellipsoid can be obtained by:
\begin{equation}
 S = 4\pi\left\{\frac{(a_1a_2)^p + (a_2a_3)^p + (a_3a_1)^p}{3}\right\}^{1/p},
\end{equation}
where $p=1.6075$ and $a_1$, $a_2$ and $a_3$ are the semi-axes of the ellipsoid. 
Using the semi-axes of the elliptic grains, the effective size of the grain can also be written as $a=(a_1 a_2 a_3)^{1/3}$. 
Hereafter, the size of a grain is referred to as the effective size of the grain.

\begin{table}
\tbl{Physical properties of graphite }{%
\begin{tabular}{cclclc} 
\hline
& \multicolumn{1}{c}{Symbol} & Unit & & Value & \\
\hline\noalign{\vskip3pt} 
 & $\rho$ & $\mathrm{g\:cm^{-3}}$ & & 2.24 \footnotemark[$*$] & \\
 & $M$ & ${\rm g\,mol^{-1}}$ & & 12.0 \footnotemark[$\dag$] & \\
 & $d$ & ${\rm \AA}$ & & 0.71 \footnotemark[$\dag$]\footnotemark[$\ddag$] & \\
 & $\eta$ & & & 0.169 & \\ [2pt] 
\hline
\end{tabular}}\label{table:physparams}
\begin{tabnote}
 \footnotemark[$*$] Draine and Li (\yearcite{Draine2001}) \\
 \footnotemark[$\dag$] \citet{ChemicalDic} \\
 \footnotemark[$\ddag$] Since the distance to the nearest carbon atom is $1.42\,{\rm \AA}$ \citep{ChemicalDic}, the radius of the carbon atom is set to half the distance.
\end{tabnote}
\end{table}

We assumed the continuous distribution of ellipsoid (CDE) for the grain shape distribution, as suggested by Bohren and Huffman (\yearcite{Bohren1983}). 
The CDE encompasses those grains with extreme shapes, such as when one of the semi-axes approaches infinity ($a_i\rightarrow\infty$). 
However, such morphologies are unrealistic, given the finite size and atom count of the grains. 
As a result, grains with $N_{\rm s}$ larger than $N_{\rm a}$ were excluded from the CDE. 
The Debye temperatures of the VSGs were calculated by averaging equation (\ref{eq:surfdebye}) with the adjusted CDE.

The total number of vibration modes in a graphite grain with a radius of $a$ is $N_{\rm m} = 3(N_{\rm a}-2)$. 
These modes consist of two types: the in-plane carbon-carbon (C-C) modes and the out-of-plane C-C modes. 
Both these modes conform to the two-dimensional Debye model. 
As per DL01, one-third of the total vibration modes correspond to the out-of-plane mode, with the remaining two-thirds attributed to the in-plane mode. 
The Debye temperatures are defined as 863 K ($\Theta_\mathrm{\infty,op}$) for the out-of-plane mode and 2504 K ($\Theta_\mathrm{\infty,ip}$) for the in-plane mode (DL01).
These values are considered for larger graphite grains. 
For smaller grains, surface effects and the CDE are factored in, resulting in revised Debye temperatures --- for instance, an out-of-plane ($\Theta_\mathrm{op}$) temperature of 770 K and an in-plane ($\Theta_\mathrm{ip}$) temperature of 2230 K for $a=1$ nm graphite grain. 

The first excitation energy of vibration modes is described by DL01. For a spherical grain:
\begin{eqnarray}
 \frac{\varepsilon_{\rm 1,op}}{k_{\rm B}} &=& 47.9 \,{\rm K} \,\left(\frac{a}{1\,{\rm nm}}\right)^{-1}, \\
 \frac{\varepsilon_{\rm 1,ip}}{k_{\rm B}} &=& 98.3 \,{\rm K} \,\left(\frac{a}{1\,{\rm nm}}\right)^{-1},
\end{eqnarray}
for the in-plane modes and out-of-plane modes, respectively. 
DL01 considered that graphite grains are spherical only when they contain 102 or more carbon atoms (corresponding to $a\gtrsim0.6$ nm). 
For the smaller grains, DL01 considered grains as planar and modified the first excitation energy of vibration modes. 
However, since the effect of such modification is negligible for our result, we used the above first excitation energy of vibration modes for all sizes down to the smallest size $a=0.35$ nm.

When the grain is in a thermal equilibrium state with temperature $T$, the thermal energy of the vibration modes of graphite is described as:
\begin{eqnarray}
 E_\mathrm{vib}(T,a) &= (N_a-2) k_{\rm B} \left[\Theta_\mathrm{op}f_{2}\left(\frac{T}{\Theta_\mathrm{op}},\,\frac{\varepsilon_\mathrm{1,op}}{k_{\rm B}T}\right) \right. \nonumber\\
 & \left.+2\Theta_\mathrm{ip}f_{2}\left(\frac{T}{\Theta_\mathrm{ip}},\,\frac{\varepsilon_\mathrm{1,ip}}{k_{\rm B}T}\right)\right], \label{eq:ene_gra}
\end{eqnarray}
\begin{equation}
 f_n(x,y) \equiv n\int_{y}^1\frac{t^ndt}{{\rm e}^{t/x}-1}.
\end{equation}
The lower bounds of the integral for function $f_n$ have been assigned a finite value that accounts for the lowest energy of the vibration mode. 
Explicit information about defining the temperature for graphite VSGs is provided in subsection \ref{sec:IETB}. 

\subsection{Thermal energy carried by free electrons \label{sec:ele}}

Referring to section \ref{sec:intro}, the forms of the VSGs are considerably non-uniform, causing the energy level distribution of their free electrons of the VSGs to differ from one to another.
It is unfeasible to detail the thermal characteristics, such as heat capacity, of individual grains.
Nevertheless, we observe the emissions and absorptions attributable to the collective VSGs. 
We need to understand not the energy level distribution of each grain, but its statistical features. 
Because VSG's shape distribution is anticipated to be highly inconsistent, the successive energy levels of free electrons around the Fermi level can be effectively represented using the random matrix theory \citep{Mehta1990}. 
This concept underpins the energy level statistics approach \citep{Kubo1962}. 

According to random matrix theory \citep{Mehta1990}, the difference in a matrix's eigenvalues follows the Wigner distribution when each matrix element is allocated random Gaussian distributed values. 
The Wigner distribution's derivation can be found in appendix \ref{App:wigner}. 
This study utilizes the Wigner distribution as a model for the interval ($\Delta$) distribution between two successive free electron energy levels near the Fermi energy. 
It describes the probability of encountering two successive energy states with an energy interval from $\Delta$ and $\Delta+d\Delta$ as:
\begin{equation}
 P_\mathrm{W}(\Delta,a)d\Delta = \frac{\pi}{2}\frac{\Delta}{\delta}\exp \left[-\frac{\pi}{4}\left(\frac{\Delta}{\delta}\right)^2\right]\frac{d\Delta}{\delta},
\end{equation}
where $\delta$ is the ensemble average of $\Delta$ among VSGs of size $a$. The $\delta$ is defined as the reciprocal of the density of states at Fermi energy level, $D(E_{\rm F})$, as $\delta = D(E_{\rm F})^{-1} = 3E_{\rm F} / (4N_{\rm e})$ where $N_{\rm e}$ is the number of free electrons given by $N_{\rm e} = N_{\rm a} = 471\,(a/1\,\mathrm{nm})^3$ for graphite grain of radius $a$. 
Therefore, $\delta$ depends on the grain size $a$.
The $D(\varepsilon)$ is the density of states of free electrons with energy $\varepsilon$, which is given by $D(\varepsilon) = 3N_{\rm e} \varepsilon^{1/2} / (4E_{\rm F}^{3/2})$. 
Since the ground state of the free electrons is the Fermi energy level, the energy of the first excitation state is not restricted by the grain size, unlike the energy of the first excitation state of the vibration modes.
Therefore, the $\Delta$s can take continuous values from $\Delta\sim0$ to about $\delta$. 

At low temperatures, where $T<\delta/k_{\rm B}$, free electrons' excitation to the second or higher excitation levels tends to be disregarded due to the minimal likelihood of such reaction happening.
We assume that the grain is in thermal equilibrium at temperature $T$.
The grain's probability of reaching the second excitation state is directly proportional to $\exp[-(\Delta_1 + \Delta_2)/(k_{\rm B}T)]$, where $\Delta_1$ and $\Delta_2$ respectively stand for the space between the ground state and first excitation state, and the space between the first and second excitation states. 
The chance of both $\Delta_1$ and $\Delta_2$ becoming equal to or smaller than $k_{\rm B}T$ is virtually non-existent when $T<\delta/k_{\rm B}$. 
Hence, we focus solely on the first excitation state and the ground state in our model for estimating the internal energy possessed by the free electrons at low temperatures. 
Under this assumption, the partition function of the free electron system is given by:
\begin{equation}
 Z_\mathrm{even}(T, \Delta) = 1 + 4\,{\rm e}^{-\beta \Delta},
\end{equation}
when the number of electrons is even, and:
\begin{equation}
 Z_\mathrm{odd}(T, \Delta) = 2\left(1 + 2{\rm e}^{-\beta\Delta}\right),
\end{equation}
when the number of electrons is odd, where $\beta\equiv(k_{\rm B}T)^{-1}$. 
The derivation of these partition functions is given in appendix \ref{App:partfunc}. 
As explained in appendix \ref{App:partfunc}, the contribution from holes is taken into account. 
The partition function of the grain is assumed to be given by the average of the two, $Z=(Z_\mathrm{even} + Z_\mathrm{odd})/2$. 
The ensemble average of thermal energy per single grain among the VSGs is given by:
\begin{equation}
 E_\mathrm{ele}(T,a) = -\frac{\partial}{\partial \beta} \left[\int^{\infty}_{0} P_{\rm W}(\Delta,a) \log Z(\Delta) d\Delta \right].\label{eq:eneele}
\end{equation}

In the high-temperature limit, the thermal energy of a grain carried by free electrons is well represented by the usual Fermi gas model \citep{Kittel1976} as:
\begin{equation}
 E_\mathrm{ele}(T,a) = \frac{\pi^2}{3}\frac{(k_{\rm B}T)^2}{\delta}. \label{eq:fermigas}
\end{equation}

\section{Method for describing thermal emission from graphite grains\label{sec:method}}

\subsection{Internal energy equivalent temperature and internal energy bound \label{sec:IETB}}

The thermalization of internal energy within dust grains occurs very swiftly \citep{Birks1970, Leger1988}. 
Consequently, each grain aligns with the law of energy equipartition.
VSGs, per \citet{Tielens2005}, are heated by stochastic absorption of photons from the interstellar radiation field (ISRF). 
However, the photon absorption interval for a VSG, exceeding a few thousand years (DL01), is significantly longer than its thermalization time scale.
Hence, VSGs can be viewed as thermally isolated systems. 
Referring to subsection \ref{sec:ele}, the degrees of freedom for the free electrons, crucial to the thermal processes in a cold graphite VSG, can effectively be considered infinite. The dust grain's temperature can thus be determined using:
\begin{equation}
 E_\mathrm{gr}(T,a) = E_\mathrm{vib}(T,a) + E_\mathrm{ele}(T,a), \label{eq:defgraintemp}
\end{equation}
where $E_\mathrm{gr}(T,a)$ is the internal energy held by the grain, $E_\mathrm{vib}(T,a)$ and $E_\mathrm{ele}(T,a)$ are the thermal energy defined by equation (\ref{eq:ene_gra}), and by both equations (\ref{eq:eneele}) and (\ref{eq:fermigas}), respectively. 
We refer to the temperature defined by equation (\ref{eq:defgraintemp}) as the internal energy equivalent temperature (IET). 
Figure \ref{fig:internalenergy} shows that the dominant component which holds thermal energy switches from vibration modes to free electrons at a temperature around $T_1=\varepsilon_{\rm 1,op}/k_{\rm B}$.
Below $T_1$, the free electrons become dominant. 
It is expected that the systematic errors introduced by using equation (\ref{eq:ene_gra}) to express the thermal energy carried by the vibration modes at temperatures around and below $T_1$ is not so significant. 
Therefore, we use the expression as the model to describe thermal energy held by the vibration modes for all temperature ranges.

The emissivity due to the thermal emission from a dust grain is given by:
\begin{eqnarray}
 j_\nu = C_{\rm abs}(\nu)S_\nu,\label{eq:thermem}
\end{eqnarray}
where $C_{\rm abs}(\nu)$ is the absorption cross-section of the grain at frequency $\nu$, and $S_{\nu}$ is the source function at frequency $\nu$.
The source function of the thermal emission from the grain is defined by:
\begin{eqnarray}
 S_{\nu}&=&\tilde{B}_{\nu}(T), \label{eq:defSourceF}
 \end{eqnarray}
where:
\begin{equation}
 \tilde{B}_\nu(T) = \left\{
 \begin{array}{cc}
 B_\nu(T) & (h\nu \leq E_\mathrm{gr}), \\
 0 & (h\nu > E_\mathrm{gr}),
 \end{array}
 \right. \label{eq:defieb}
\end{equation}
where $h$ is the Planck constant, $B_{\nu}(T)$ is the Planck function at a frequency $\nu$ and $T$ is the IET. 
The upper limit of the frequency arises from the constraint that the energy of emitted photons cannot exceed the internal energy of the grain. 
This modification was previously introduced by \citet{Bron2014} to calculate the probability distribution of VSG temperature. 
We refer to this upper limit as the internal energy bound (IEB)

\subsection{Base model}\label{sec:basemodel}

\begin{figure}[ht!]
 \begin{center}
 \includegraphics[width=80mm]{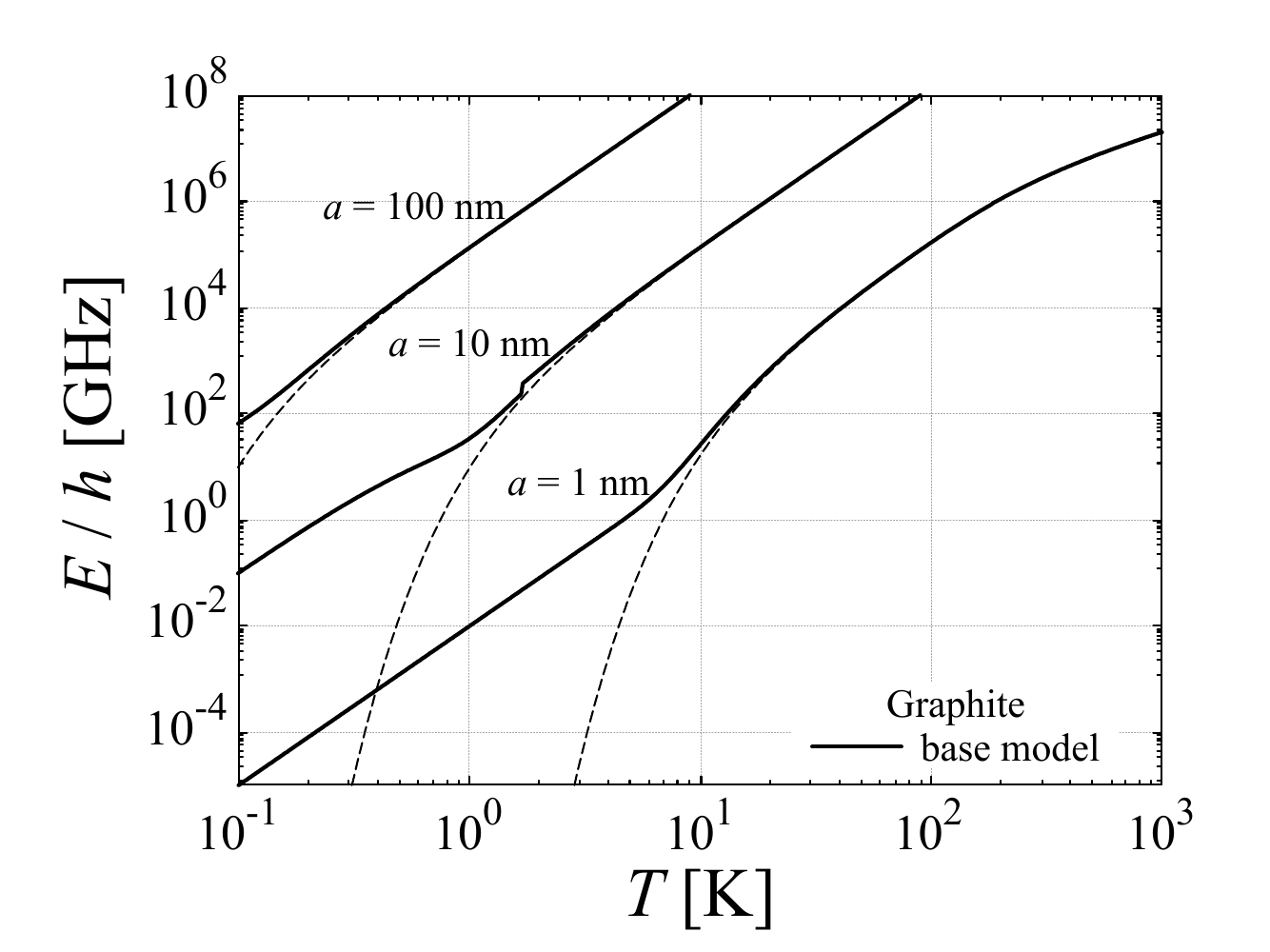}
 \end{center}
 % \plotone{figure/Energy_gra_02.pdf}
 \caption{Internal energy of graphite grains of three different sizes ($a=1,10,100$ nm). The vertical axis represents the internal energy in terms of GHz. The dashed lines represent the thermal energy held by the vibration modes. The solid lines show the sum contributions from the vibration modes and the free electrons.}
 \label{fig:internalenergy}
\end{figure}

The thermal properties of each graphite grain's free electrons are assumed to be outlined by models explained in subsection \ref{sec:ele}. We switch the formula describing the thermal energy held by free electrons regarding whether $E_{\rm ele}$ is smaller than $\delta$ or not.
Specifically, equation (\ref{eq:eneele}) applies when $E_\mathrm{ele}$ is less than $\delta$, and equation (\ref{eq:fermigas}) is used when $E_\mathrm{ele}$ exceeds $\delta$.
While the thermal energy model for free electrons introduced in subsection \ref{sec:ele} is more relevant for VSGs and is not suited for larger grains, we have expanded its application for simplicity. 
The model considering the surface effect, mentioned in subsection \ref{sec:surf}, is adopted for the internal energy stored by the lattice vibration modes. 
The internal energy of graphite grains, graphed as a function of temperature, can be seen in figure \ref{fig:internalenergy}. 
It demonstrates the principal carrier of the grain's internal energy --- being the lattice vibration modes for grains larger than $a=10$ nm --- within the temperature range of our study (section \ref{sec:result}). 
Consequently, the contribution from free electrons ceases automatically for larger grains, which means there is no need for concern when applying the VSG's free electron thermal energy model to them.

When $T < T_1$, the energy from vibration modes decreases exponentially. 
As the temperature increases, these modes of vibration start to align with the Debye model. 
At very low temperatures, free electrons begin to make significant contributions to the energy.
However, in nano-sized grains, the influence of free electrons becomes dominant below 10 K. 
Conversely, big grains of $a=100$ nm size adhere well to the Debye model, even at these low temperatures. 
Emissivity, which arises from thermal emissions from a dust grain, is dealt with using the models mentioned in subsection \ref{sec:IETB}. 
This model is henceforth referred to as the base model. 

We adopt the absorption cross-section of the grain, $C_{\rm abs}(\nu)$, as reported by \citet{Laor1993} for all frequencies. This is the model of spherical crystal grains.
We extrapolate with $C_\mathrm{abs}\propto \nu^2$ for frequencies below $300$ GHz. 
This study does not focus on the polarization emission from the grains, so we neglect the shape dependence of the absorption efficiency in this paper. 
According to Bohren and Huffman (\yearcite{Bohren1983}), any differences in the wavelength range extending beyond the far-infrared are not significant. 

\subsection{Reference model} \label{sec:reference}

The impact of the mesoscopic properties is clarified through probability distributions of temperature (IET) and SEDs that have been calculated for a reference model. 
This model defines the energy of the vibration mode's first excitation level as zero ($\varepsilon_1=0$) and assumes a continuous distribution for the energy levels of vibration modes. 
The Debye temperature, with no surface effects, is used, and the free electrons' contribution is excluded.
The standard Planck function is employed as the source function for thermal emission, and the IET is used to represent the dust grain's temperature.

\subsection{Monte-Carlo simulation code to follow the thermal history of a dust grain \label{sec:MCcode}}

A Monte-Carlo simulation code has been created to track a dust grain's thermal history.
This code enables us to monitor the grain's stochastic heating process due to absorption of photons from the surrounding ISRF and the CMB, as per the Monte-Carlo methodology \citep{Draine1985}. 
The code also treats radiative cooling as a continuous cooling process, which happens due to the thermal emission from the grain. 

The calculation for stochastic heating processes proceeds as follows: 
The frequency of a photon absorbed by the grain is sampled from the range between $\nu_{\rm min}=3$ GHz and $\nu_{\rm max}= 7.5\times10^6$ GHz. 
Although Draine and Anderson (\yearcite{Draine1985}) modeled the grain heating due to absorption of photons with frequencies smaller than 300 GHz as a continuous process, we treated the photon absorption at any frequency as a stochastic process.
For the stochastic absorption computation, we assess the probability density, $p(\nu,a,G_0)$, representing the odds that the grain absorbs a photon of frequency $\nu$ per unit time per unit frequency. 
Rather than randomly sampling $\nu$ between $\nu_{\rm min}$ and $\nu_{\rm max}$, we avoid potential sampling bias among diverse frequency bands by sampling a random value, $x$, between 0 and 1.
This sampled $x$ value is then converted to $\nu$ using the equation $\nu=\nu_{\rm min}\left(\nu_{\rm max}/\nu_{\rm min}\right)^x$. The absorption probability density is then represented as follows:
\begin{equation}
 p(\nu,a,G_0) = \frac{C_\mathrm{abs}(\nu,a)cu_\nu(G_0)}{h} \ln\left(\frac{\nu_{\rm max}}{\nu_{\rm min}}\right), \label{eq:absprob}
\end{equation}
where $c$ is the speed of light and $u_\nu$ is the energy density of the ambient radiation field which is described as:
\begin{equation}
 u_\nu(G_0) = G_0u_\nu^\mathrm{ISRF} + \frac{4\pi}{c}B_\nu(T_\mathrm{CMB}).
\end{equation}
Here, $u_\nu^\mathrm{ISRF}$ is the ISRF at the solar neighborhood described in \citet{Mathis1983}, $G_0$ is the scale factor with $G_0=1$ representing that of the solar vicinity, and $T_\mathrm{CMB}=2.725\mathrm{K}$ is the temperature of the CMB \citep{Fixsen2009}. 
The average time interval between photon absorptions, $\tau_{\rm abs}(a,G_0)$, is defined as $\tau_{\rm abs}(a,G_0)^{-1} \equiv \int^{\nu_\mathrm{max}}_{\nu_\mathrm{min}}p(\nu,a,G_0)d\nu$. For $a=1$ nm graphite, $\tau_\mathrm{abs} = 6.57\times10^5$ seconds when $G_0=1$.

We utilized a continuous radiative cooling model for the thermal emission from the grain.
The evolution of the IET during the time interval between a photon absorption and the successive photon absorption is calculated by:
\begin{eqnarray}
 \frac{dT}{dt} = -\frac{1}{C_V(T,a)}\int^{\infty}_{0}C_{\rm abs}(\nu,a)4\pi \tilde{B}_\nu(T)d\nu, \label{eq:cooling}
\end{eqnarray}
where $C_V(T,a)\equiv dE_{\rm gr}(T,a)/dT$ is the heat capacity of the grain. 
We define a cooling timescale, $\tau_{\rm cool}(T,a)$, by $\tau_\mathrm{cool}(T,a)\equiv E_{\rm gr}(T,a) / \dot{E}_{\rm gr}(T,a)$ where $\dot{E}_{\rm gr}(T,a)\equiv dE_{\rm gr}(T,a)/dt = C_V(T,a)\,dT/dt$. 

The time-variations of dust grains due to stochastic heating and continuous cooling are shown in figure \ref{fig:thermhis} in appendix \ref{App:tdist}.
For each simulation, we continued until 50,000 absorption events had taken place \citep{Draine1985}.

We obtained the probability distribution function of IET, $dP/dT$, of the simulated grain, where $dP/dT\times \Delta T$ provides the probability of finding the simulated grain in the temperature range between $T$ and $T+\Delta T$.
A detailed explanation of our method for extracting $dP/dT$ is provided in appendix \ref{App:tdist}. 
The time-averaged emissivity of thermal emission from a single grain of size $a$ (for a fixed $G_0$) at a frequency $\nu$ is calculated by:
\begin{equation}
 i_\nu(a,G_0) = \int^{\infty}_{0}C_\mathrm{abs}(\nu,a)4\pi\tilde{B}_\nu(T)\frac{dP(T,a,G_0)}{dT} dT. \label{eq:singleemit}
\end{equation}

The total SED of interstellar graphite dust grains is comprised of emission from numerous dust grains, each of a particular size, $a$.
To compute this, we initially reinterpret $dP/dT\times \Delta T$ under the ergodic hypothesis as the probability of randomly selecting a grain that maintains a temperature between $T$ and $ T+\Delta T$ within the ensemble \citep{Purcell1976, Draine1985}. 
The variety of the interstellar graphite dust grain sizes can be ascertained through a specified size distribution function, $d n(a)/da$ where $d n(a)/da \times da $ provides the number density of grains with size between $a$ and $ a+da$ \citep{Weingartner2001}. 
The total intensity of the graphite dust thermal emission, denoted as $I_\nu$, is determined by averaging equation (\ref{eq:singleemit}) within this size distribution.
\begin{equation}
 \frac{I_\nu(G_0)}{N_{\rm H}} = \frac{1}{4\pi}\frac{1}{n_{\rm H}}\int^{a_{\rm max}}_{a_{\rm min}} \frac{dn(a)}{da}i_\nu(a,G_0)da, \label{eq:sed}
\end{equation}
where $n_{\rm H}$ and $N_{\rm H}$ are the number density and the column density of the hydrogen atoms along the line of sight, respectively. 
The $a_{\rm min}$ and $a_{\rm max}$ are the minimum and the maximum size of the grains. 
Equation (\ref{eq:sed}) represents the total SED of the thermal emission from the interstellar graphite dust grains predicted by the base model. 

\subsection{Comparison with DustEM code}\label{sec:dustem}

We evaluated the SED of thermal emissions from interstellar graphite dust grains obtained using our code in comparison to results achieved with DustEM \citep{Compiegne2011}, a well-known tool for calculating interstellar dust SEDs.
In DustEM, the static temperature distribution of these dust grains is computed employing the iterative method \citep{Desert1986} and modeling relies on the reference model.
The SEDs were derived for the reference model featuring $G_0=1$. 
Our study uses the size distribution from Weingartner and Draine (\yearcite{Weingartner2001}), and a total-to-selective extinction ratio ($R_V$) of 3.1 \citep{Tielens2005}. Additionally, we assumed a carbon abundance of $b_{\rm C}=6\times10^{-5}$ within VSGs sized between $a_{\rm min}=0.35$ nm and $a_{\rm max} = 1\, {\rm \mu m}$. Here, all the carbonaceous dust grains are assumed to be graphite.

\begin{figure}[ht!]
 \begin{center}
 \includegraphics[width=80mm]{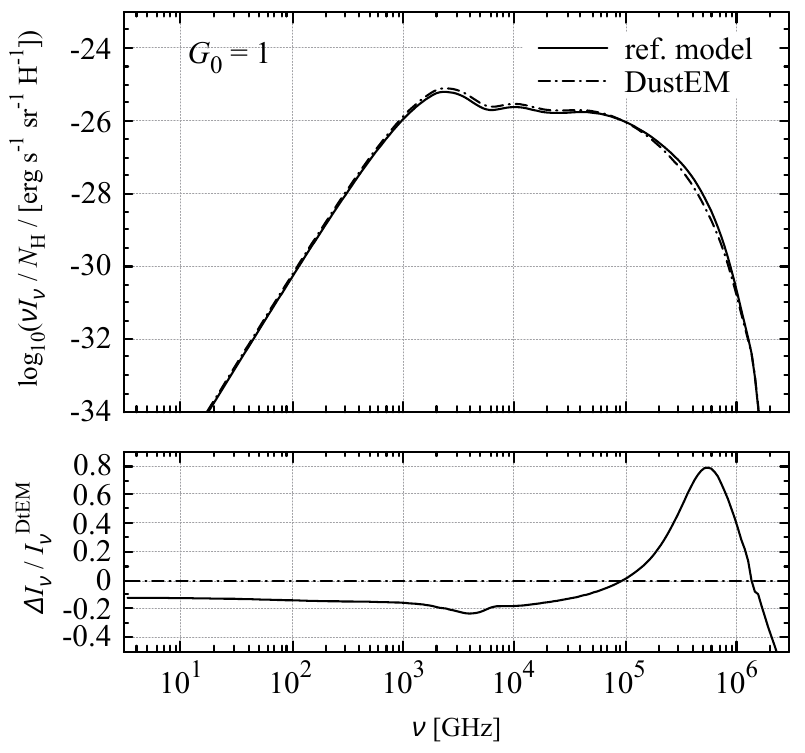}
 \end{center}
 \caption{Top panel: The SED of thermal emission from graphite dust grains calculated with our code (solid line) and DustEM (dash-dotted line). The SEDs are calculated for the reference model with $G_0=1$. The size distribution is derived from Weingartner and Draine (\yearcite{Weingartner2001}) with $R_V=3.1$ and $b_{\rm C}=6\times10^{-5}$ (see text). The size range is between $a_{\rm min}=0.35\,{\rm nm}$ and $a_{\rm max} = 1\, {\rm \mu m}$. Bottom panel: the difference between the two results relative to the SED calculated by the DustEM. }
 \label{fig:dustem}
\end{figure}

In figure \ref{fig:dustem}'s top panel, the SED determined by our code is on display alongside the results acquired from DustEM. 
The bottom panel illustrates the residual $\Delta I_{\nu}\equiv I_\nu - I_{\nu}^{\rm DtEM}$, where $I_{\nu}$ and $I_{\nu}^{\rm DtEM}$ represent the intensity of the thermal emission per hydrogen atom computed by our code and DustEM, respectively. 
The two SEDs demonstrate substantial agreement across all frequency ranges. 
Especially notable is their concurrence below $10^4$ GHz.
This reaffirms the suitability of our code's application for predicting SEDs in sub-millimeter and millimeter wavebands. 

\section{Results \label{sec:result}}

\subsection{Results when the heat source is only the CMB \label{sec:CMBonly}}

\begin{figure*}[ht!]
 \begin{center}
 \includegraphics[width=160mm]{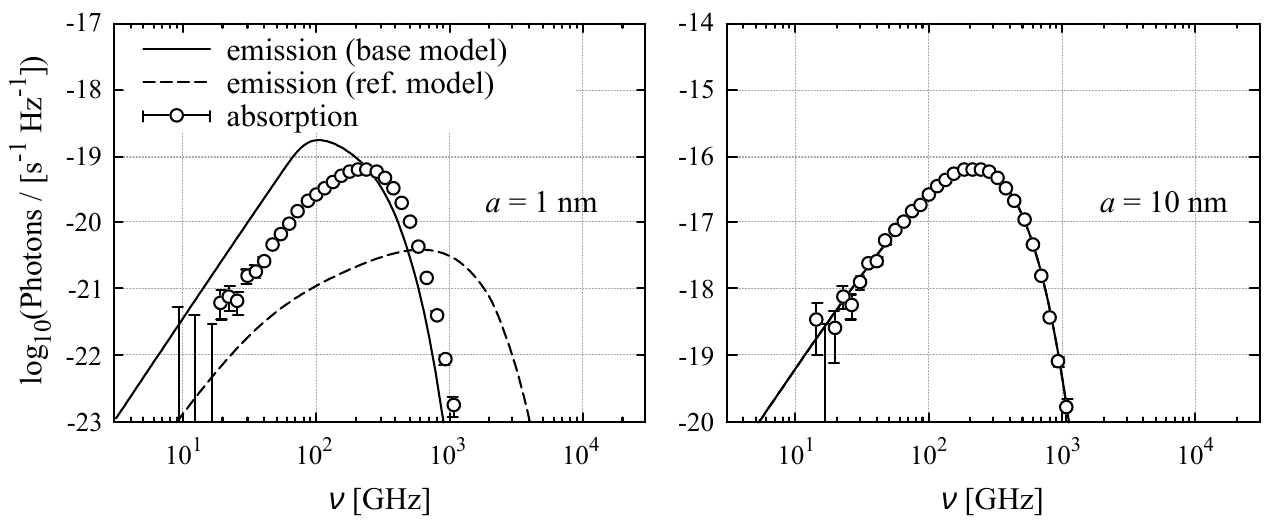}
 \end{center}
 \caption{The absorption and emission rates of a single graphite dust grain of $a=1\,{\rm nm}$ (left panel) and $a=10\,{\rm nm}$ (right panel). The heat source is solely the CMB. Vertical axes are the number of photons absorbed or emitted by the grain per unit time per unit frequency. Circles are results of absorbed photons with Poisson errors assigned as the error bars for the absorption rate. They are identical for both the base model and the reference model. The solid lines show results of emitted photons when the base model is adopted. The dashed lines show results of emitted photons when the reference model is adopted.}
 \label{fig:specu0}
\end{figure*}

\begin{figure*}[ht!]
 \begin{center}
 \includegraphics[width=160mm]{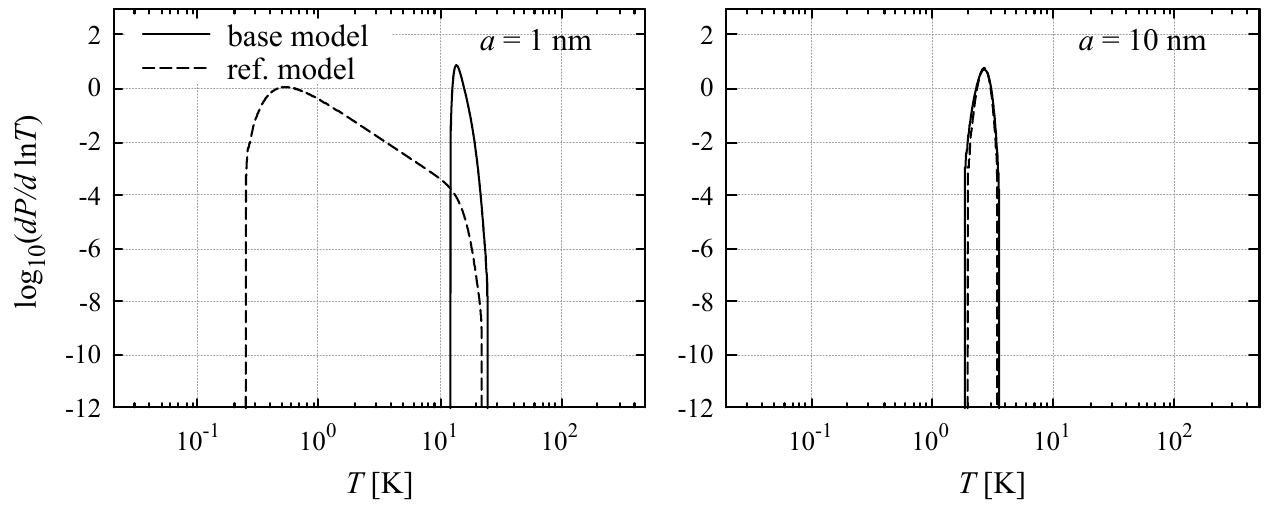}
 \end{center}
 \caption{The probability distribution function of temperature (IET) of graphite dust grains calculated with the base model (solid lines) and with the reference model (dashed lines) when grain sizes are $a=1\,{\rm nm}$ (left panel) and $a=10\,{\rm nm}$ (right panel). The heat source is solely the CMB.}
 \label{fig:tdistu0}
\end{figure*}

We examined results using only CMB photons as the heat source to better understand the base model's diagnostics.
We simulated the IET probability distribution for large grains with $a=100$ nm, immersed in the CMB. We found that the distribution of IET is sharply focused at the CMB temperature, meaning the radiative equilibrium is maintained with high accuracy. 

Figure \ref{fig:specu0} illustrates the photon absorption and emission rates per unit frequency per unit time for a single grain with sizes $a=1$ nm and $a=10$ nm. 
These rates indicate the likelihood of a grain either absorbing or emitting a photon within a specific time interval ($dt$) and frequency range ($\nu$ to $\nu +d\nu$).
This is calculated by multiplying $dt$ and $d\nu$ by the relevant absorption and emission rates at a frequency $\nu$.
Here is how we determine the absorption rate:
The frequency range is divided into bins with a fixed logarithmic interval. 
The number of absorbed photons is tallied for each bin during one simulation run.
The absorption rate is then calculated by dividing the counted number of absorbed photons by the duration time and the frequency interval.
The duration of each simulation is approximately $10^{10}$ seconds for $a=1$ nm grain and approximately $10^7$ seconds for $a=10$ nm grain. 
The emission rate is $i_{\nu}/(h\nu)$.
The probability distributions of IET for both cases are shown in figure \ref{fig:tdistu0}. 
For comparison, the emission rate in the case of the reference model is overlaid.

Figure \ref{fig:specu0} illustrates nearly identical absorption and emission rates for $a=10$ nm grain, with the differences between the base model and reference model being minimal. 
Conversely, for an $a=1$ nm grain, the reference model's emission rate significantly deviates from its absorption rate, even though the absorption rates for both models remain the same.
Meanwhile, the base model displays emission and absorption rates with a similar distribution, with the former showing a slight shift towards a lower frequency.

The IET distributions are depicted in figure \ref{fig:tdistu0}. 
For $a=10$ nm grain, the results predicted using the base model and the reference model are nearly identical, with a sharp peak at the CMB temperature of 2.725 K.
Conversely, for $a=1$ nm grain, the probability distribution of IET deduced from the reference model spans a broad range from approximately $0.2$ K to $20$ K, with the majority of grains remaining below 1 K. 
In the case of the base model, the distribution has a sharp peak at 13 K.

The sharp 13 K peak of the IET distribution is due to a strong IET dependence of the cooling timescale. 
At around 13 K, the internal energy of a VSG is $E_{\rm gr}(T,a)\ll k_{\rm B}T$. 
The photon frequency emitted by the grain is limited to $\nu<E_{\rm gr}(T,a)/h\ll k_{\rm B}T/h$, due to the IEB.
Therefore, the source function is approximated by $2k_{\rm B}T\nu^2/c^2$. 
At this frequency range, the absorption cross section can be written as $C_{\rm abs}(\nu,a)=C_0(\nu/\nu_0)^2$, where $C_0=5\times10^{-20}\,{\rm cm^2}$ and $\nu_0=10^3$ GHz for $a=1$ nm graphite \citep{Laor1993}.
Thus, the emission rate is calculated as:
\begin{eqnarray}
    \dot{E}_{\rm gr}(T,a) &\simeq& \int^{E_{\rm gr}(T,a)/h}_{0}C_{0}\left(\frac{\nu}{\nu_0}\right)^2\frac{8\pi\nu^2}{c^2}k_{\rm B}T d\nu \nonumber \\
    &\propto& T\cdot E_{\rm gr}(T,a)^5.
\end{eqnarray}
The internal energy of the grain can be written as $E_{\rm gr}(T,a) \simeq E_0(T/10\;{\rm K})^3$ at $T\sim 10$ K, where $E_0/h = 100$ GHz for $a=1$ nm grain (figure \ref{fig:internalenergy}). Consequently, the cooling timescale depends on IET as $\tau_{\rm cool}(T,a) \simeq 2\times10^{8}(T/10\,{\rm K})^{-13}$ seconds. 
The IET which satisfies $\tau_{\rm abs}(a,G_0)=\tau_{\rm cool}(T,a)$ is $T_{\rm eq}\sim15$ K.
Since the IET dependence of $\tau_{\rm cool}(T,a)$ is extremely strong, the grain's IET concentrates at around $T_{\rm eq}$. This is the reason why the IET distribution has a sharp peak at around 13 K.

In the reference model, the frequency of the peak position of the emission rate is approximately $10^3$ GHz, corresponding to the peak position of the gray body spectrum with a temperature around $13$ K. 
This occurs because the IEB is not set in the reference model, allowing dust grains to emit photons beyond the IEB.
Consequently, dust grains can emit photons that exceed their internal energy, which is physically unacceptable. 
While this does not result in a negative temperature due to the adoption of the continuous cooling model, it leads to an overestimation of the cooling rate. 
The majority of the grains remain below 1 K. This is the artifact of the unphysical modeling of the reference model.

The base model also radiates around 13 K, but because of the IEB, the frequency of the emitted photon is limited to below $E_\mathrm{gr}(T,a)/h$. 
This frequency is significantly smaller than the peak frequency of the gray body spectrum with $T\sim 13$ K.
Since the emissivity below $E_\mathrm{gr}(T,a)/h$ depends on the frequency as $\propto \nu^4$, most of the thermal energy is emitted at the frequency $\nu=E_\mathrm{gr}(T,a)/h$. 
Hence, a graphite VSG emits a photon at nearly the same frequency as the absorbed photon, explaining why the emission rate closely aligns with the absorption rate. 
Additionally, non-zero contributions from the frequencies lower than that of the absorbed photons cause a slight shift of the emission rate towards the lower frequency side.

\subsection{Results when heat sources are CMB and starlight}\label{sec:CMBandstarlight}

\begin{figure*}[ht!]
 \begin{center}
 \includegraphics[scale=0.55]{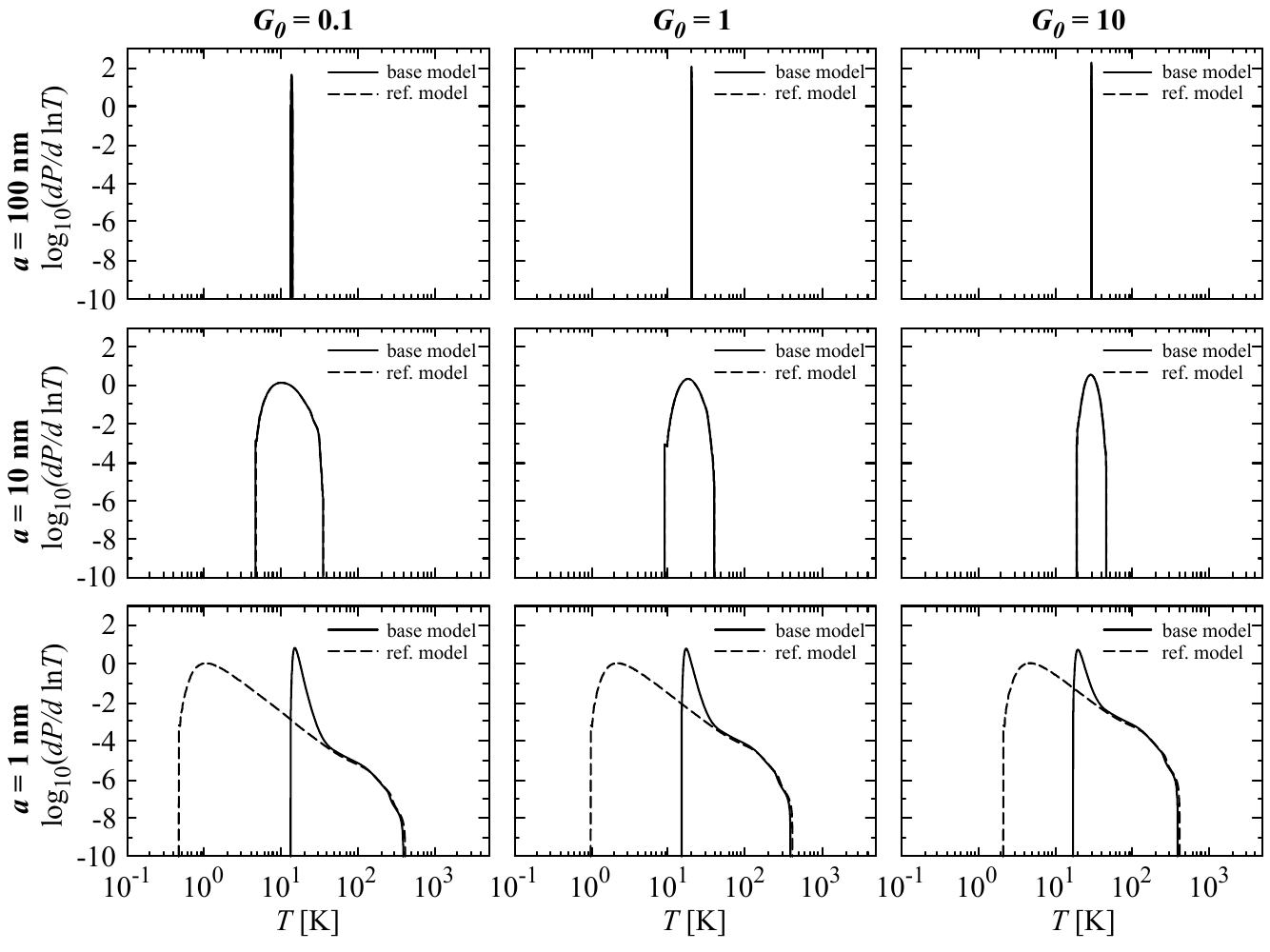}
 \end{center}
 \caption{The probability distribution of temperature (IET) of graphite grains of different sizes ($a=1,10,100\,{\rm nm}$) and different intensities of radiation field ($G_0=0.1, 1, 10$). The solid lines represent the probability distributions of the base model, and the dashed lines represent the reference model. The discrepancies between the two models are negligible for the big grains ($a=10,100\,{\rm nm}$).}
 \label{fig:tdist}
\end{figure*}

\begin{figure*}[ht!]
 \begin{center}
 \includegraphics[scale=0.55]{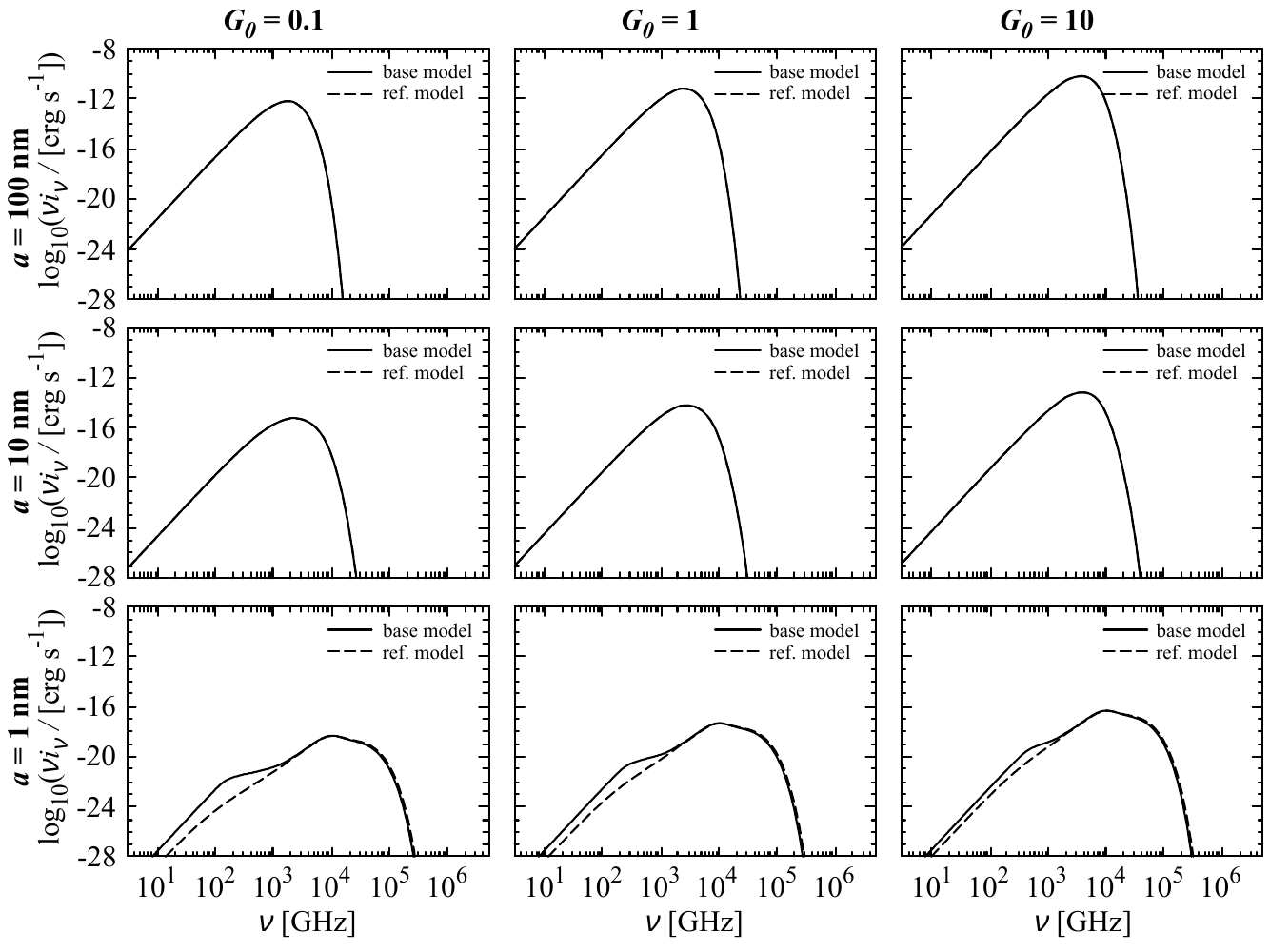}
 \end{center}
 \caption{The dust thermal SED of graphite dust grains with three different sizes ($a=1, 10, 100\,{\rm nm}$) and different intensities of ISRF ($G_0=0.1, 1, 10$). The SEDs of each panel were calculated using the probability distribution of IET shown in figure \ref{fig:tdist}. The solid lines represent the SED of the base model, and the dashed lines represent the reference model. The discrepancies between the two models are negligible for the big grains ($a=10,100\,{\rm nm}$).} 
 \label{fig:sed}
\end{figure*}

The probability distributions of IET for a single grain and the SEDs of thermal emission per single grain when the grain is exposed to both the ISRF and the CMB are examined using our code.
Figure \ref{fig:tdist} displays the distributions of IET for graphite dust grains of three different sizes: $a=1, 10, 100$ nm, and at three different intensities of the ISRF ($G_0 = 0.1, 1, 10$). 
The results obtained from the base model are compared with those from the reference model. 
In the top panels of figure \ref{fig:tdist}, it is observed that the IET of large grains ($a=100$ nm) remain nearly constant for each $G_0$. 
Specifically, IETs are recorded as 12 K for $G_0=0.1$, 20 K for $G_0=1$, and 30 K for $G_0=10$, consistent with IETs obtained under radiative equilibrium \citep{Tielens2005}. 
The middle panels illustrate that for medium-sized grains ($a=10$ nm), the probability distributions of IET peaked around radiative equilibrium temperatures but are broadened.
Predictions from the base and reference models are almost identical for larger and medium-sized grains. 

The bottom panels show the IET distributions of graphite VSGs ($a=1$ nm).
The base model predicts that the IET distributions are confined to above approximately $10$ K. 
The physical reason for the sharp peaks around $T\sim 20$ K is the same as seen in figure \ref{fig:tdistu0}.
While the peak positions of the IET distributions do not significantly vary among different $G_0$ values, they do shift slightly towards higher temperatures with increasing $G_0$.
Features around $10$ K in the base model's distribution are absent in the reference model's distributions, with most grains remaining around a few K in the reference model. 
This discrepancy originates from the same source as seen in figure \ref{fig:tdistu0}. 
Both models exhibit long tails in the high-temperature side of the IET probability distributions, extending up to a few hundred K, which is not depicted in figure \ref{fig:tdistu0}.
These high-temperature tails are attributed to the absorption of ISRF photons with a typical frequency $\nu_{\rm ISRF} \sim 2\times10^5$ GHz \citep{Tielens2005}.
The similarity in the probability distributions of IET in the high-temperature region predicted by both models arises because the IEB frequency becomes very high at high IETs, mitigating the effect of the IEB. 

Slight deviations in the IET distributions at several hundred K, reflect differences in the Debye temperatures. 
The Debye temperature of the base model is smaller than that of the reference model due to the surface effect and the irregular grain shapes. 
Consequently, even with the same total internal energy, the grain exhibits a smaller IET with the base model.

Thermal emission from a single graphite dust grain is represented by the time-averaged SEDs shown in figure \ref{fig:sed}.
These SEDs are derived from the IET probability distributions illustrated in figure \ref{fig:tdist}.
Figure \ref{fig:sed}'s top and middle panels display the SEDs of a larger ($a=100$ nm) and medium-sized ($a=10$ nm) grain, respectively. 
The SEDs, which show peaks in the far-infrared wavelength correspond well with the gray body spectra with a few 10 K.
For both the large and medium-sized grains, their SEDs are nearly identical.

The SEDs for the VSGs ($a=1$ nm) are displayed in the bottom panels of figure \ref{fig:sed}.
When compared to the reference model results, the SEDs that result from the base model demonstrate excess emissions in the millimeter to sub-millimeter wavebands. 
As $G_0$ increases, both the frequency of excess emission and deviations from the reference model predictions decrease, shifting from millimeter to sub-millimeter wavebands. 

Both models' SEDs exhibit a plateau extending from peak positions to higher frequencies up to the mid-infrared. 
This trait arises from the VSGs briefly reaching several hundred K in temperature \citep{Seki1978, Draine1984, Draine1985}. 
With an increase in $G_0$, the highest frequencies of the plateaus shift slightly towards higher frequencies.
The minor discrepancies between the high-frequency region ($\nu>10^4$ GHz) SEDs, predicted by both models, can be traced back to the variations in IET distribution at several hundred K.

\section{Discussion \label{sec:discussion}}

\subsection{The total SED of graphite dust grains}
\label{sec:flattening}

\begin{figure*}[ht!]
 \begin{center}
 \includegraphics[width=160mm]{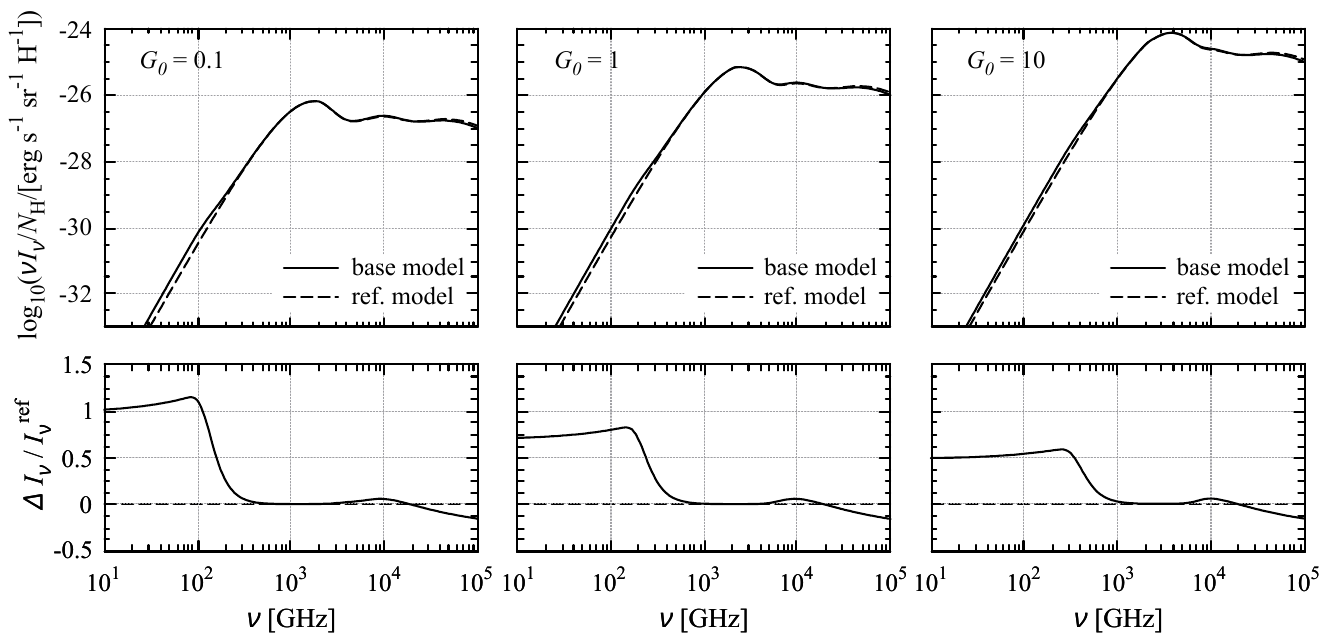}
 \end{center}
 \caption{Top panels: SEDs of thermal emission from interstellar graphite dust grains for various intensities of the ISRF ($G_0=0.1, 1, 10$). SEDs are calculated by taking the size distribution weighted average. The size distribution is adopted from \citet{Weingartner2001} with $R_V=3.1$ and $b_C=6\times10^{-5}$ for grain sizes between $a_{\rm min}=0.35\,{\rm nm}$ and $a_{\rm max} = 1\,{\rm \mu m}$. The solid and dashed lines represent the SEDs of the base model and the reference model, respectively. Bottom panels: the residual $\Delta I_\nu\equiv (I_\nu - I^{\rm ref}_\nu)$ where $I^{\rm ref}_\nu$ is the thermal emission calculated using the reference model.}
 \label{fig:totsed}
\end{figure*}

We have demonstrated how mesoscopic physics influence the total SED of the thermal emission from interstellar graphite dust grains when majority of very small carbonaceous grains are graphite.
We used the size distribution of carbonaceous dust grains provided by Weingartner and Draine (\yearcite{Weingartner2001}).
Figure \ref{fig:totsed} depicts the total SEDs of thermal emission from interstellar graphite dust grains per unit hydrogen atom as defined by equation (\ref{eq:sed}). 
The size range from $a=0.35$ nm to a maximum of $a=1\,{\rm \mu m}$ was divided into 50 bins with an equal logarithmic interval. 
The total SEDs were calculated by superposing SEDs obtained for each size bin.
The differences between the predictions of the base model and the reference model are seen in the lower panels in figure \ref{fig:totsed}, displaying the SED residuals.
For $G_0=1$, there is prominent excess emission below 500 GHz.
The intensity of excess emissions predicted through the base model is comparable to that of the reference model predictions for the millimeter waveband.
Figure \ref{fig:totsed} reveals that the excess sub-millimeter emission from the graphite VSGs is still noticeable even when the emissions from large grains are overlaid, yielding a flatter apparent slope in the sub-millimeter wavelength of the SED.
Both model's SEDs demonstrate a plateau beyond the peak frequency, extending from $10^3$ GHz frequency up to $10^5$ GHz. 
This presents the mid-infrared excess emission relative to a gray body spectrum with a temperature of a few 10 K. 
The mid-infrared excess emissions detected in the base and the reference models are almost identical. 
This excess emissions in both wavebands stems from the VSGs.

\subsection{Limitations of using internal energy equivalent temperature} \label{sec:limitationofIET}

It is anticipated that when grains are only heated by the CMB, the radiative processes of the dust grains will not alter the CMB spectrum.
This is because the interaction between the CMB photons and the dust grains remain in a stationary state, and the CMB possesses a blackbody spectrum. 
Our findings indicate that the emission and absorption rates of large grains are consistent with each other.

The base model's forecasted emission rate for VSGs roughly aligns with the absorption rate.
While this appears to satisfy the previously stated condition, a minor discrepancy exists.
Moreover, the IET of the VSGs are observed to be higher than the CMB temperature (figure \ref{fig:tdistu0}). 
This signifies a challenge when defining the thermodynamic temperature of the VSGs.
Consequently, our emission model needs validation against an alternative emission model, one that does not rely on the concept of ``temperature" (DL01). 

\subsection{Extending the base model to other dust species}\label{sec:extendingmodel}

This subsection discusses extending our graphite model to other dust species, including PAHs and silicates.
The base model can extend to dust types possessing free electrons, such as PAHs.
PAHs are thought to contain free electrons as their molecular structure, akin to graphene, contains $\pi$ electrons \citep{Atkins2017}. Consequently, it is plausible these $\pi$ electrons behave as free electrons within PAHs. 
\citet{Hoang2017} posited that the $\pi$ electrons in PAHs mimic the role of free electrons, enabling the PAHs to interact with the local magnetic field. 
This interaction should yield observable, polarized emission from the PAHs.
However, the absorption cross section of PAH is different from that of graphite \citep{Draine2007}, necessitating a calculation of PAH SEDs with the above-mentioned mesoscopic physics taken into account.

\citet{Hoang2016} have proposed the potential existence of nanosilicates as a source of anomalous microwave emission (AME). 
Distinctly, unlike graphite grains or PAHs, silicates lack free electrons, necessitating an alternate model to adequately represent the thermal properties of these purported nanosilicates.
At present, the only observational evidence suggesting the existence of nanosilicates, is the AME itself.
Furthermore, the nanosilicate model is one of several proposed explanations for the origin of AME, a matter that continues to be contested \citep{Draine1998a, Draine1999, Jones2009, Nashimoto2020a, Nashimoto2020b}.

\subsection{Quantum interference effect of free electrons}\label{sec:interference}

In addition to the above-mentioned mesoscopic physics, the quantum interference of free electrons' wave function becomes vital when modeling the physical properties of VSGs. 
This is primarily because the de Broglie wavelength of electron clouds linked to each graphite atom surpasses the VSG's size.
The de Broglie wavelength is calculated using the formula $\lambda_{\rm dB}=h/p$, where $p$ symbolizes the electron cloud's momentum.

Assuming that the electron cloud's mass encircling each graphite atom is quadruple that of an electron, the electron cloud's momentum, at temperature $T$, can be estimated as $p\sim (8m_{\rm e}k_{\rm B}T)^{1/2}$. 
Consequently, the de Broglie wavelength results as $\lambda_{\rm dB}\sim20\,{\rm nm}\,(T/10\,{\mathrm K})^{1/2}$, which is equal to or greater than the grain size.

It is imperative to investigate how this quantum effect impacts astronomical observables.
Furthermore, it is intriguing to consider whether this presents a fresh prospect for examining the quantum effect manifesting in nano-scale metallic grains through astronomical observations. 

\section{Conclusion}\label{sec:conclusion}

We have developed a model to illustrate the thermal properties of cosmic very small graphite dust grains (graphite VSGs), recognizing that these grains, which consist of 100 to 10,000 atoms, embody mesoscopic-scale materials. 
For the first time, we have factored in the impact of free electrons on the thermal properties of the graphite VSGs. 
To achieve this, we used the energy level statistics, method first introduced by Kubo (1962) to explain the thermal characteristics of a nano-scale metal grains.

Our findings reveal that, by acknowledging the irregularity of grain shapes, the grains can maintain a continuous distribution of energy levels for free electrons in relation to the Fermi level from zero.
The degree of freedom of these free electrons, which play a crucial role in the thermal processes within a cold graphite VSG, can be perceived as effectively infinite.
This realization allows us to introduce the concept of ``temperature" when discussing the thermal properties of a cold graphite VSG.
To further illustrate this point, we introduced the internal energy equivalent temperature (IET)(figure \ref{fig:internalenergy}).
Additionally, we factored in the reduction of the Debye temperature due to the surface effect. 

We propose a model for the emissivity of thermal emission from the graphite VSGs using the IET. 
This model considers the frequency's upper limit, which a VSG can emit, based on its internal energy.
The VSGs' low-temperature energy content significantly affects the emitted spectra due to their minimal nature.

We developed a Monte-Carlo simulation code to determine the probability distributions of IET and the SEDs of thermal emissions from a dust grain.
Our code incorporates stochastic heating from ambient photon absorption and continuous radiative cooling.
We validated this code by comparing the SEDs it produced with those derived from DustEM, confirming its accuracy (figure \ref{fig:dustem}).

We presented findings where CMB solely heats graphite dust grains. 
The absorption and emission rates of large grains exhibited near-perfect agreement.
For graphite VSGs, the emission rate mimicked the absorption rate's pattern, albeit with a minor discrepancy (figure \ref{fig:specu0}).
This difference suggests that accurately defining the thermodynamic temperature of VSGs may pose challenges.
Therefore, an update to the emission model is necessary. 

We have analyzed the IET distributions and SEDs of graphite dust grains in interstellar environments. 
Our findings reveal that the graphite VSGs are responsible for the excess mid-infrared emission and also produce a significant amount of energy at sub-millimeter and millimeter wavelengths (figure \ref{fig:sed}).

Very small cosmic dust grains could serve as a natural laboratory for investigating the physics of mesoscopic systems.

\begin{ack}
KA acknowledges support from the Graduate Program on Physics for the Universe (GP-PU), Tohoku University. 
KA and MH acknowledge support from JSPS KAKENHI grant number JPJSBP120219943 Bilateral Joint Research Project.
MN acknowledges support from JSPS KAKENHI grant No. JP22J00388.
This work was supported by JST SPRING, Grant Number JPMJSP2114, JSPS Core-to-Core Program JPJSCCA20200003, and MEXT KAKENHI grant Nos. JP18H05539, JP20KK0065.
We thank the anonymous referee for helpful comments.

\end{ack}

\appendix

\section{Wigner distribution} \label{App:wigner}
Assume a system that can be represented by $2\times2$ real symmetric Hamiltonian matrix $\textsf{H}$.
\begin{eqnarray}
 \textsf{H} &=& \left(
 \begin{array}{cc}
 h_{11} & h_{12} \\
 h_{21} & h_{22}
 \end{array}
 \right).
\end{eqnarray}
For simplicity, we write $h_{11} = 2h_1,\,h_{22} = 2h_{2},\, h_{12} = h_{21}=2h$. The $h_1$, $h_2$ and $h$ are assumed to independently follow the Gaussian with mean 0 and variance $(2\alpha)^{-1/2}$ with $\alpha$ being a constant. 
Then, the probability distribution function of $(h_1,h_2,h)$ can be written as
\begin{eqnarray}
 P(h_1, h_2, h) \propto {\rm e}^{-\alpha((2h_1)^2 + (2h_2)^2 + 2\cdot(2h)^2)}
 \label{eq:probh}
\end{eqnarray}

The following procedure can diagonalize the Hamiltonian:
\begin{eqnarray}
 \textsf{H} = \textsf{O}\left(
 \begin{array}{cc}
 \lambda_{+} & 0 \\
 0 & \lambda_{-}
 \end{array}
 \right)\textsf{O}^{\rm T},
\end{eqnarray}
\begin{equation}
    \textsf{O} = \left(
 \begin{array}{cc}
 \sin\theta & \cos\theta \\
 \cos\theta & -\sin\theta
 \end{array}
 \right),
\end{equation}
\begin{equation}
    \lambda_{\pm} = h_1 + h_2 \pm h_0.
\end{equation}
Here, $\lambda_{\pm}$ is an eigenvalue of the Hamiltonian, $\cos\theta \equiv 2h/[2h_0(h_1 - h_2 + h_0)]^{1/2}$, $\sin\theta \equiv (h_1- h_2 + h_0)/[2h_0(h_1 - h_2 + h_0)]^{1/2}$ for $0 \leq \theta \leq \pi/2$, and $h_0\equiv[(h_1 - h_2)^2 + 4h^2]^{1/2}$. 
 The Jacobian of the transformation from $(h_1, h_2, h)$ to $(\lambda_+,\lambda_-,\theta)$ can be written as $J=-(\lambda_+-\lambda_-)$. 
 Therefore, the distribution function of $(\lambda_+,\lambda_-,\theta)$ can be written as:
 \begin{eqnarray}
 P(\lambda_+, \lambda_-, \theta) &=& P(h_1, h_2, h)\,||J|| \nonumber\\
 &\propto& (\lambda_+ - \lambda_-) {\rm e}^{-\alpha(\lambda_+{}^2 + \lambda_-{}^2)}.
 \end{eqnarray}

We write the energy interval of the two levels as $\Delta\equiv\lambda_+ - \lambda_-$. 
We also define $\mathcal{E}\equiv\lambda_+ + \lambda_-$. 
The distribution function of $(\Delta,\mathcal{E},\theta)$ can be written as:
\begin{eqnarray}
 P(\Delta, \varepsilon, \theta) &\propto \Delta\,{\rm e}^{-\frac{\alpha}{2}\Delta^2}\,{\rm e}^{-\frac{\alpha}{2}\mathcal{E}^2}.
\end{eqnarray}

The energy state of the system is now specified by the three variables $(\Delta,\mathcal{E},\theta)$, and these are all independent from each other. 
For now, we are only interested in the distribution of $\Delta$, so $\mathcal{E}$ and $\theta$ can be treated as constants. 
The normalization condition defines the proportionality constant $\int^{\infty}_{0} P(\Delta)\,d\Delta = 1$ and $\int^{\infty}_{0} \Delta \cdot P(\Delta)\,d\Delta = \delta$ where $\delta$ is the expectation value of $\Delta$. 
The distribution function of $\Delta$ follows the Wigner distribution which is described as:
\begin{eqnarray}
 P(\Delta) = \frac{\pi}{2}\left(\frac{\Delta}{\delta}\right)\,{\rm e}^{-\frac{\pi}{4}\left(\frac{\Delta}{\delta}\right)^2}.
\end{eqnarray}

\section{Partition function of the free electron system}\label{App:partfunc}

Here, we derive the partition function of the free electron system.
The energy levels in bulk crystals are highly degenerate when the grain shapes are symmetric. 
In mesoscopic systems, the degeneracy of the energy levels of free electrons is broken since irregularities of the grain shapes become prominent.
Energy of each level is represented by $\varepsilon_i$ ($i=\cdots,-2, -1, 0, 1, 2, \cdots$) where $\varepsilon_0$ is the Fermi level. 
We set $\varepsilon_0=0$.

The heat capacity of the grains is primarily determined by the electrons, which are excited to energy states above the Fermi level, as well as the holes excited to energy states below the Fermi level. 
Holes are generated at positions where electrons become absent following the excitation. 
When the number of free electrons is small (less than a few hundred), whether the total number of electrons is even or odd affects the physical properties of the grain, necessitating separate consideration of these cases.
Additionally, since the free electron system serves as a crucial carrier of internal energy for the VSGs only when the grain's temperature significantly drops below the average interval of adjacent energy levels, contributions from second or higher excitation levels can be disregarded.

\begin{figure}[ht!]
 \begin{center}
 \includegraphics[scale=0.45]{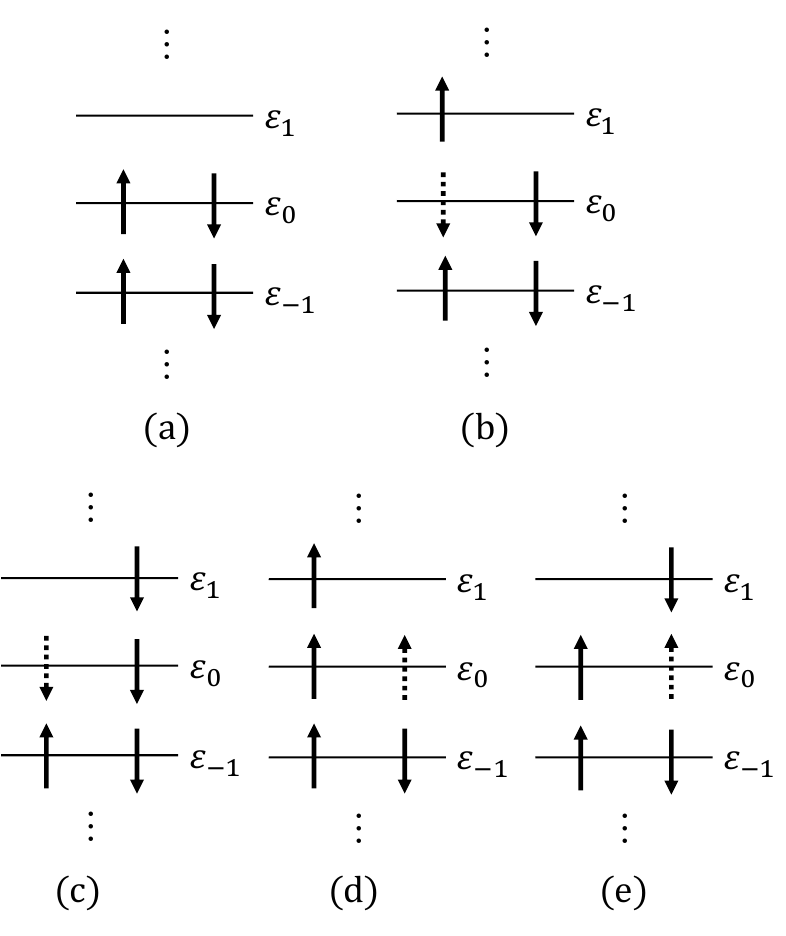}
 \end{center}
 \caption{Energy states of free electrons when the total number of electrons is even. The solid arrows represent electrons, and the dotted arrows represent holes. The direction of the arrows represents the spin direction.}
 \label{fig:statenumeven}
\end{figure}

\begin{figure}[ht!]
 \begin{center}
 \includegraphics[scale=0.45]{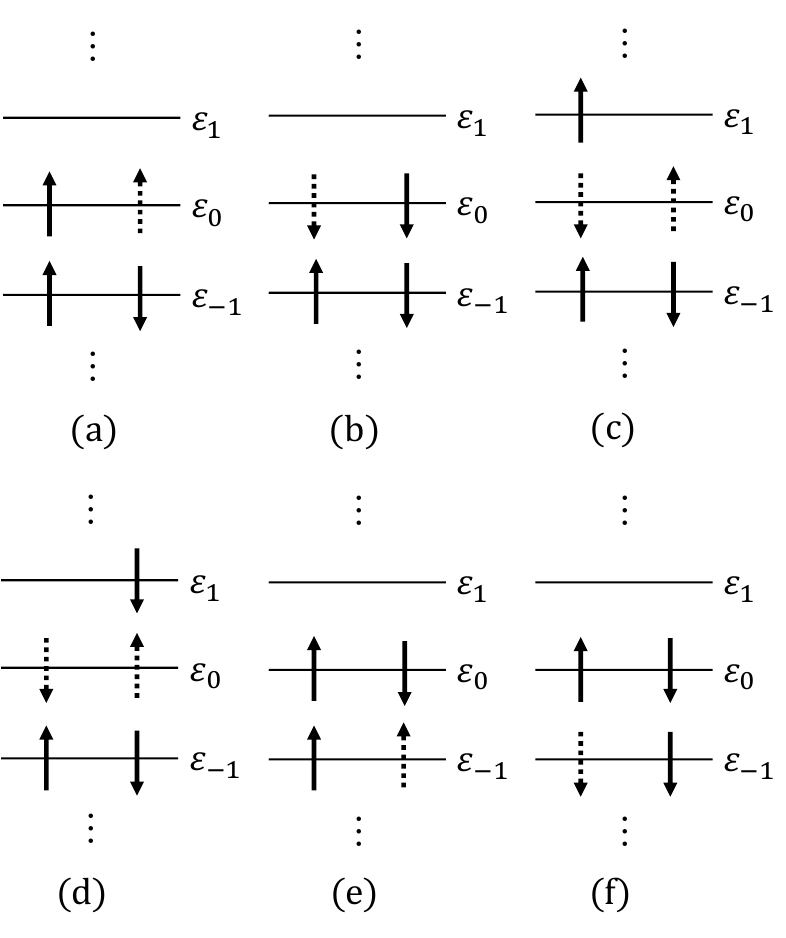}
 \end{center}
 \caption{Same as figure \ref{fig:statenumeven} but for the case when the total number of electrons is odd.}
 \label{fig:statenumodd}
\end{figure}

The potential configurations of electrons and holes in the energy levels across the Fermi energy are illustrated in figure \ref{fig:statenumeven} and \ref{fig:statenumodd} for cases when the number of electrons is even and odd, respectively. 
In figure \ref{fig:statenumeven}(a), the ground state is depicted when the number of electrons is even, while figure \ref{fig:statenumeven}(b) - \ref{fig:statenumeven}(e) portray excited states. 
These diagrams are integrated into the subsequent partition function as shown below.
\begin{eqnarray}
 Z_{\rm even} = 1 + 4 {\rm e}^{-\beta\Delta} \label{eq:partfunceven}, 
\end{eqnarray}
where $\Delta = \varepsilon_1 - \varepsilon_0$.

Similar diagrams but for odd number electrons are shown in figure \ref{fig:statenumodd}. 
Figure \ref{fig:statenumodd}(a) and \ref{fig:statenumodd}(b) represent the ground states, and figure \ref{fig:statenumodd}(c) - \ref{fig:statenumodd}(f) are the excited states. 
The partition function becomes:
\begin{eqnarray}
 Z_{\rm odd} = 2 \left(1 + 2{\rm e}^{-\beta\Delta}\right). \label{eq:partfuncodd}
\end{eqnarray}

\section{Extracting the probability distribution function of IET
from the thermal history of a dust grain} \label{App:tdist}

\begin{figure}[ht!]
 \begin{center}
 \includegraphics[width=80mm]{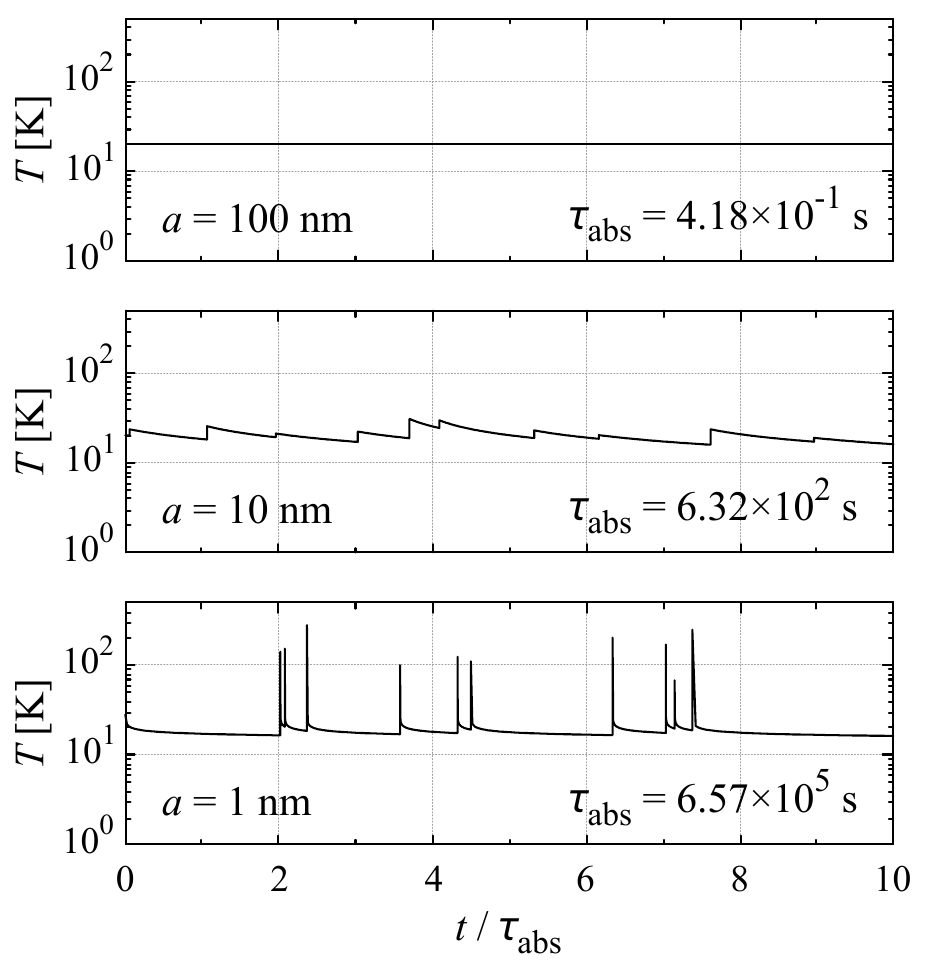}
 \end{center}
 \caption{The time-variations of graphite grain temperature (IET) of three different sizes ($a=1, 10, 100$ nm) calculated using the base model with $G_0=1$. The horizontal axis is the time divided by $\tau_{\rm abs}$ which is the average time interval between photon absorptions described in subsection \ref{sec:MCcode}.}
 \label{fig:thermhis}
\end{figure}

The thermal history of graphite grains for $G_0=1$ is demonstrated in figure \ref{fig:thermhis}. The horizontal axis represents time divided by the average time interval of photon absorptions described in subsection \ref{sec:MCcode}. For $a=1$ nm and 10 nm grains, each spike represents the abrupt rise in IET due to absorption of a single photon. 
For $a=100$ nm grain, the IET is almost constant at its radiative equilibrium temperature \citep{Tielens2005} because the change in the IET caused by absorption of a single photon is negligible.

The probability distribution function of IET, $dP/dT$, provides the probability of finding a dust grain in the temperature range between $T$ and $T+\Delta T$ as $dP/dT\times \Delta T$ at a certain time. 
We calculate the total duration that the grain stays in the IET range from $T$ to $T'(>T)$, where $T'-T=\Delta T$, during each simulation by the following method. 
We assume that the temperature of the grain rises abruptly to a new IET when the grain absorbs a photon \citep{Birks1970, Leger1988}. 
A duration that a grain cools down from $T'$ to $T$ is described by:
\begin{equation}
 \Delta t = \int_{T}^{T'} \left|\frac{dT}{dt}\right|^{-1}dT, \label{eq:residencetime}
 \end{equation}
where $dT/dt$ is described by equation (\ref{eq:cooling}). 
We count how many times the grain passes through the temperature range from $T'$ to $T$ due to the cooling. 
When the count is $x$, the total duration that the grain stays in the IET range from $T$ to $T'$ becomes $x\Delta t$. 
Although the duration that the grain stays in the temperature range becomes shorter than that given by equation (\ref{eq:residencetime}) when the absorption of a photon during the grain stays in the temperature range, we neglect this effect to calculate the total duration. 
Because $\Delta T$ is set to be small enough, this introduces negligible error in the obtained probability distribution function of IET. 
A $T_{\rm min}$ is set to be 0.9 times the minimum value of the IET that the grain reached during each simulation, and a $T_{\rm max}$ is set to be 1.1 times the maximum value of the IET that the grain reached during the simulation.
The range from $T_{\rm min}$ to $T_{\rm max}$ is divided into $N$ bins with a fixed logarithmic interval. 
We have set $N=1000$. 
The probability of finding a grain in the $i$th bin is 
deduced by $P_i=x_i \Delta t_i/\sum (x_i \Delta t_i)$ where $x_i$ is the count 
that the grain stays in the $i$th IET bin and $\Delta t_i$ is the duration that the grain passes the $i$th temperature bin calculated by equation (\ref{eq:residencetime}). 
The probability distribution function of IET is obtained by dividing $P_i$ by the temperature interval.

\end{document}